\documentclass[aps,prx,twocolumn,amsmath,amssymb,superscriptaddress,letterpaper]{revtex4}
\usepackage{times}
\usepackage{amsfonts}
\usepackage{mathrsfs}
\usepackage[version=3]{mhchem}
\usepackage{graphicx}
\usepackage{dcolumn}
\usepackage{bm}
\usepackage{color}

\usepackage[colorlinks,bookmarks=false,citecolor=blue,linkcolor=red,urlcolor=blue]{hyperref}
\def\be{\begin{equation}}       \def\ee{\end{equation}}
\def\bea{\begin{eqnarray}}      \def\eea{\end{eqnarray}}

\begin{document}
\title{Effective model and pairing tendency in bilayer Ni-based superconductor La$_3$Ni$_2$O$_7$}
\author{Yuhao Gu}
\thanks{These authors equally contributed to the work.}
\affiliation{Beijing National Laboratory for Condensed Matter Physics and Institute of Physics, Chinese Academy of Sciences, Beijing 100190, China}

\author{Congcong Le}
\thanks{These authors equally contributed to the work.}
\affiliation{ 
RIKEN Interdisciplinary Theoretical and Mathematical Sciences (iTHEMS), Wako, Saitama 351-0198, Japan}

\author{Zhesen Yang}
\affiliation{Department of Physics, Xiamen University, Xiamen 361005, Fujian Province, China}

 \author{Xianxin Wu}\email{xxwu@itp.ac.cn}
 \affiliation{CAS Key Laboratory of Theoretical Physics, Institute of Theoretical Physics,
Chinese Academy of Sciences, Beijing 100190, China}

\author{Jiangping Hu}\email{jphu@iphy.ac.cn}
\affiliation{Beijing National Laboratory for Condensed Matter Physics and Institute of Physics, Chinese Academy of Sciences, Beijing 100190, China}
\affiliation{Kavli Institute for Theoretical Sciences, University of Chinese Academy of Sciences, Beijing 100190, China}
\affiliation{ New Cornerstone Science Laboratory, Beijing 100190, China}
\begin{abstract}
Since the discovery of cuprate, the origin of high-T$_c$ superconductivity has been an outstanding puzzle. Recently, high-T$_c$ superconductivity was observed in a bilayer nickelate La$_3$Ni$_2$O$_7$ under pressure, whose structure hosts the apical oxygen between two layers, distinct from multi-layer cuprates. Motivated by this discovery, we investigate its electronic structure using first-principle calculations and superconducting instabilities from both weak-coupling and strong-coupling perspective. Based on the first-principle band structures, we construct a bilayer two-orbital model on a square lattice, consisting of $d_{x^2-y^2}$ and $d_{z^2}$ orbitals, which accurately captures the low-energy electronic properties.  Within this model, we study pairing instability using both functional renormalization group approach and multi-orbital t-J model. An $s_{\pm}$-wave pairing with sign-reversal gaps on different Fermi surfaces is revealed, reminiscent of  iron based superconductors. The Ni-$d_{z^2}$ orbital and its associated interlayer and intralayer exchange couplings are found to be crucial for the high-T$_c$ superconductivity.  Our study provides valuable insights into unique nature of electronic structure and superconductivity in La$_3$Ni$_2$O$_7$ and contributes to the understanding of unconventional superconductors.

\end{abstract}

\maketitle

Since the landmark discovery of high-T$_c$ superconductivity in cuprates\cite{Bednorz1986}, extensive research efforts have been dedicated to  experimentally exploring noncopper-based compounds with similar crystal and electronic structure and theoretically elucidating the underlying mechanism~\cite{maeno1994,takada2003,Lee2006,PRX.5.041012,Keimer2015}. The $d^9$ configuration of Cu$^{2+}$ in an octahedral crystal field is believed to be pivotal and the corresponding electronic structure can usually be described by the one-band Hubbard model. Owing to nickel being the neighbor of copper in the periodic table and Ni$^{1+}$ hosting the same $d^9$ electron count as Cu$^{2+}$, nickelates were identified as novel candidates for achieving high T$_c$ superconductivity. Despite tremendous efforts in nickelates, superconductivity has only been realized until very recently in so-called “infinite-layer" nickelates (Sr,Nd)NiO$_2$ thin films on a substrate ~\cite{Li2019,Osada2020,Pan2022,wang2022,ding_critical_2023}. Interestingly, superconductivity emerges around 5-15 K without a magnetic order. Similar to $d^9$ cuprates, the low-energy states are dominated by the Ni $d_{x^2-y^2}$ orbital but the contribution of Nd $d$ orbitals introduce a self-doping in the Ni $d_{x^2-y^2}$ orbital~\cite{PhysRevLett.125.077003,Wu2020,GMZhang2020,gu2020,Nomura_2022}. The T$_c$ of the “infinite-layer" nickelates can be enhanced to be over 30 K with external pressure~\cite{wang2022}. Still, their maximum T$_c$ values have consistently fallen below the widely recognized  McMillan limit. 

A recent breakthrough in the field of high-T$_c$ superconductivity has emerged with the discovery of the bilayer perovskite bulk nickelate La$_3$Ni$_2$O$_7$, which demonstrates an extraordinary high superconducting T$_c=80$ K under high-pressure conditions~\cite{sun2023}. With increasing pressure, the system experiences a structural transition from the Amam phase to the more symmetric Fmmm phase. Superconductivity appears in the latter phase for a pressure larger than 14 Gpa. In sharp contrast to cuprates, the nominal valence here is $d^{7.5}$ in Ni$^{2.5+}$ and the apical oxygens between layers are preserved. Near the Fermi level, the electronic structures are dominantly contributed by Ni $d_{x^2-y^2}$ and $d_{z^2}$ orbitals. The strong $\sigma$ bonding in $d_{z^2}$ orbitals between two layers is suggested to be crucial for superconductivity~\cite{sun2023}. Therefore, the observation of high-T$_c$ superconductivity in the La$_3$Ni$_2$O$_7$ challenges the understanding of high-T$_c$ mechanism. This reignites longstanding questions in the field of unconventional superconductivity, particularly in relation to the cuprates and related materials \cite{Lee2006,Keimer2015}: (i) Whether $d^9$ configuration plays a vital role in high-T$_c$ superconductivity; (ii) The absence of observed magnetism in La$_3$Ni$_2$O$_7$ prompts the investigation of whether long-range antiferromagnetism is a key component in high-T$_c$ superconductivity; (iii) How does the d$_{z^2}$ orbital influence high-T$_c$ superconductivity.

To help address these questions, we have studied the electronic structures of La$_3$Ni$_2$O$_7$ through density functional theory (DFT) calculations and explored its superconductivity from repulsive interactions adopting both weak-coupling and strong-coupling approaches. Based on the first-principle band structures, we construct a bilayer two-orbital model on a square lattice, consisting of $d_{x^2-y^2}$ and $d_{z^2}$ orbitals, which captures the essential low-energy electronic structures. Neglecting the high-energy $d_{z^2}$ antibonding state, we obtain  a half-filled three-band model. 
There are three Fermi surfaces, with the central electron pocket attributed to a mixture of $d_{x^2-y^2}$ and $d_{z^2}$ orbitals, while the other two hole pockets around the corners are attributed to $d_{x^2-y^2}$ and $d_{z^2}$ orbitals, respectively. With the effective model, we study pairing instability using both functional renormalization group (FRG) approach and multi-orbital t-J model. An $s_{\pm}$-wave pairing with sign-reversal gaps on different Fermi surfaces is revealed, analogous to iron based superconductors~\cite{iron-based_2008,RMP.83.1589,Hirschfeld_2011}. The effect of orthorhombic distortion and three-dimensionality of electronic structures is discussed, together with possible experimental implications.

\begin{figure}
\centerline{\includegraphics[width=0.5\textwidth]{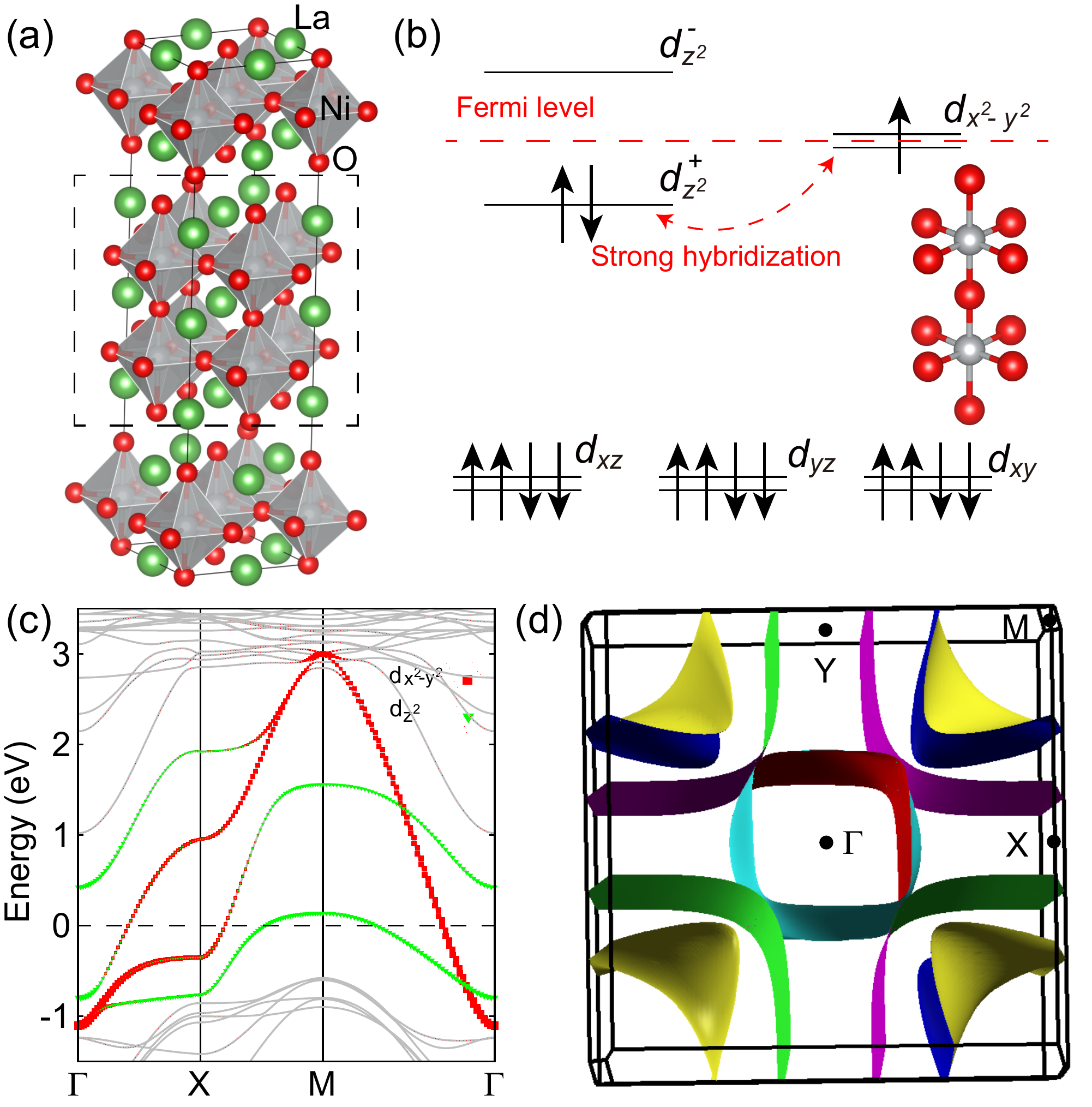}}
\caption{(color online) (a) Crystal structure of \ce{La3Ni2O7} in the high-pressure phase. (b) The illustration of the two vertex-sharing NiO$_6$ octahedra complex, the corresponding energy splitting of d orbitals, and the electronic configuration of d orbitals for two Ni$^{2.5+}$. DFT band structure (c) and Fermi surfaces (d) in the primitive cell. The orbital characters near the Fermi level are represented by different colors. \label{fig1}}
\end{figure}

{\it Electronic structures }. We start with the crystal and electronic structures of La$_3$Ni$_2$O$_7$. The high-pressure phase of La$_3$Ni$_2$O$_7$ in Fig.\ref{fig1}(a), a bilayer Ni-based superconductor, exhibits a structural arrangement characterized by the presence of NiO$_2$ bilayers, which are effectively separated by LaO layers. This intriguing configuration stems from the coordinated arrangement of two NiO$_6$ octahedra, featuring the sharing of apical oxygen atoms, as visually represented in Fig.\ref{fig1}(b). Remarkablely, these apical-oxygen-sharing NiO$_6$ octahedra form an NiO$_2$ bilayer, exhibiting an approximately square lattice structure. Here, the apical oxygen between the layers is in contrast to multi-layer cuprates.  The $e_g$ and $t_{2g}$ orbitals are well separated in energy by the octahedral crystal field, as demonstrated in Fig.\ref{fig1}(b). Furthermore, the formation of bonding and antibonding in the d$_{z^2}$ orbitals between two layer is evident from the analysis of orbital orientation. Considering the chemical valence of Ni atoms in La$_3$Ni$_2$O$_7$ as Ni$^{2.5+}$, the corresponding d-orbital  configuration is $d^{7.5}$, and thus the Fermi level is expected to  reside  between the bonding and antibonding states, as shown in Fig.\ref{fig1}(b).

 Fig.\ref{fig1}(c) displays the band structure along high-symmetry paths in the primitive cell, in agreement with Ref.\cite{luo2023bilayer}. Consistent with the analysis of crystal field splitting, near the Fermi level, the valence and conduction bands are predominantly attributed to $d_{x^2-y^2}$ orbital, bonding state $d^{+}_{z^2}$ and antibonding state d$^{-}_{z^2}$. Moreover, the band structure reveals the presence of both electron-like and hole-like Fermi surfaces in the primitive Brillouin zone. An electron-like pocket is located around the $\Gamma$ point, whereas two hole pockets manifest around the M point, as shown in Fig.\ref{fig1}(d). Notably, the two hole pockets primarily originate from the bonding state $d^{+}_{z^2}$ and the $d_{x^2-y^2}$ orbital (interlayer antibonding state), respectively. While, the electron-like pocket is attributed to the strong mixture of d$^{+}_{z^2}$ and d$_{x^2-y^2}$ orbitals (interlayer bonding state). From the Fermi surfaces, we deduce that the system is relatively two-dimensional with moderate $k_z$ dispersion and othorhobic distortion has negligible effect on the electronic structures. The Fermiology exhibits some similarities to the theoretically proposed nickelate La$_2$Ni$_2$Se$_2$O$_3$\cite{Scibull2018}.  Except enhanced in-plane bandwidth and energy splitting between $d^{+}_{z^2}$ and $d^{-}_{z^2}$ in the high-pressure phase,  the difference in electronic structures between the ambient-pressure and high-pressure phases is not significant (see supplementary material (SM)).

\begin{figure}
\centerline{\includegraphics[width=0.5\textwidth]{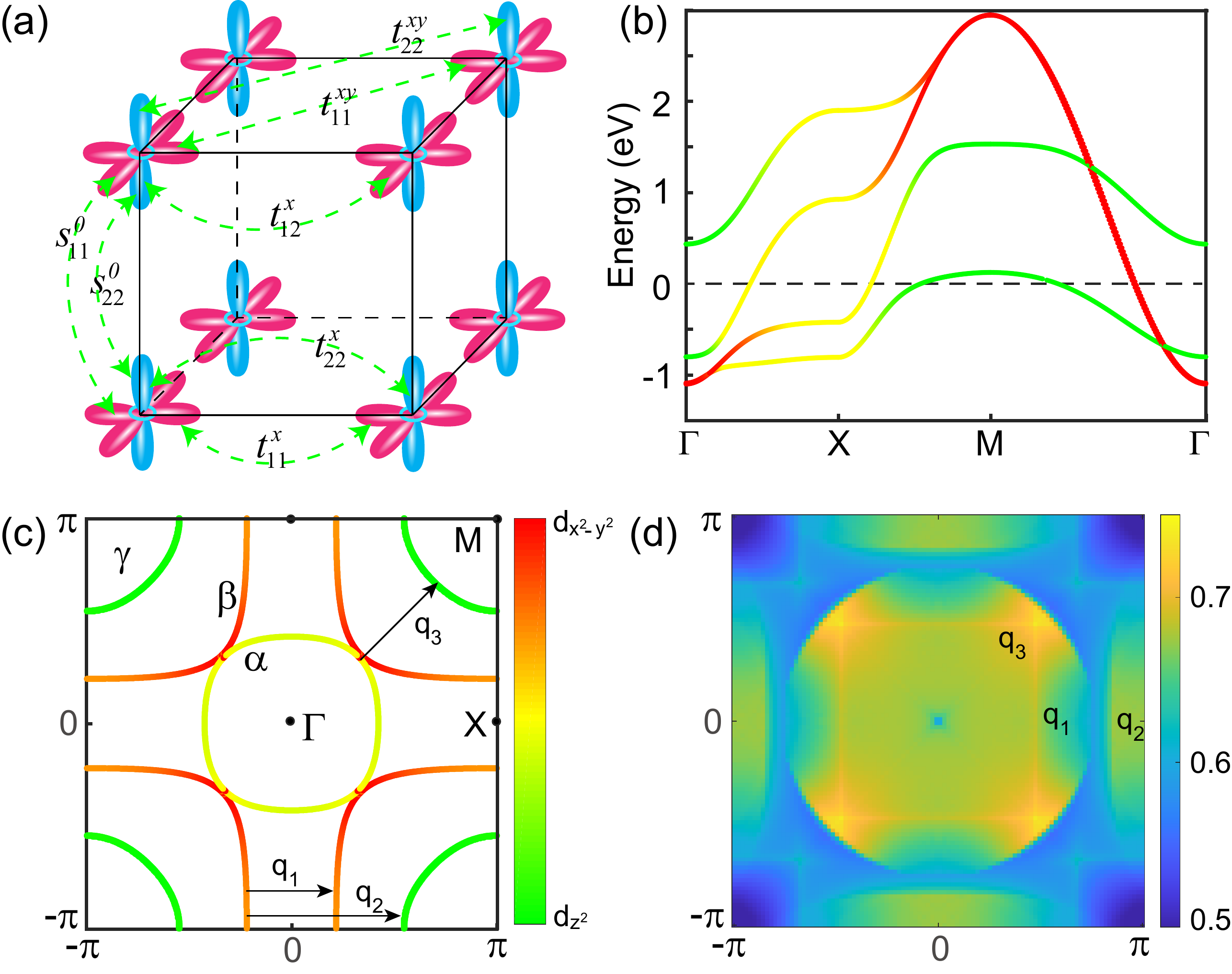}}
\caption{(color online). 
(a) The schematic of main hopping parameters in the bilayer two-orbital model. Orbital-resolved  band structure (b) and Fermi surfaces (c) from the tight-binding model with a  electron filling $n=3$. The orbital contributions are represented using different colors: d$_{x^2-y^2}$(red) and d$_{z^2}$(green). (d) Maximum eigenvalues of the susceptibility matrix $\chi_0(\bm{q})$ in the 2D Brillouin zone. \label{fig2}}
\end{figure}

{\it Bilayer two-orbital model}. Based on the aforementioned analysis, we find the $d_{x^2-y^2}$ and $d_{z^2}$ orbitals play a prominent role near the Fermi level, while the deviation from the tetragonal lattice due to distortion remains relatively weak. Consequently, a bilayer two-orbital model on a square lattice, as depicted in Fig.\ref{fig2} (a), can be employed to effectively describe the low-energy electronic structure of La$_3$Ni$_2$O$_7$ at high pressure. We introduce the operator $\psi_{\mathbf{k} \sigma}^{\dagger}=\left[c_{t x^2 \sigma}^{\dagger}(\mathbf{k}), c_{t z^2 \sigma}^{\dagger}(\mathbf{k}), c_{b x^2 \sigma}^{\dagger}(\mathbf{k}), c_{b z^2 \sigma}^{\dagger}(\mathbf{k})\right]$, where $c_{\eta \alpha \sigma}^{\dagger}(\mathbf{k})$ represents the Fermionic creation operator with $\eta$, $\alpha$ and $\sigma$ being the layer, orbital and spin indices, respectively. The layer index $\eta = t, b$ denotes the top and bottom layer and the orbital index $\alpha = 1,2$ represents the Ni d$_{x^2-y^2}$ for 1 and the Ni d$_{z^2}$ for 2. The tight-binding (TB) Hamiltonian can be written as
\begin{eqnarray}
\mathcal{H}_{\mathrm{TB}}=\sum_{\mathbf{k} \sigma} \psi_{\mathbf{k} \sigma}^{\dagger}[h(\mathbf{k})-\mu] \psi_{\mathbf{k} \sigma}.
\end{eqnarray}
The corresponding matrix elements of $h(\mathbf{k})$ and hopping parameters derived from a downfolding of the DFT bands onto four localized Wanniner functions are given in the SM. With those parameters, the obtained band fits are provided in SM and both band dispersion and orbital characters (Fig.\ref{fig2}(b)) reach a good agreement between DFT and the TB results. Along the $\Gamma-\text{X}$ path, $d_{x^2-y^2}$ and $d_{z^2}$ orbital exhibit prominent hybridization due to the strong inter-orbital hopping between the nearest-neighbor sites. Along the $\Gamma-\text{M}$ path, the $d_{x^2-y^2}$ bands are almost degenerate due to its decoupling from the $d_{z^2}$ orbitals and partial cancellation between interlayer couplings. The strong orbital mixture near the Fermi level is in sharp contrast to the bilayer cuprates and infinite nickecates~\cite{PhysRevB.89.224505,PhysRevLett.125.077003,Wu2020}. This model can be further simplified by neglecting the high-energy antibonding state $d^-_{z^2}$ and we obtain a half-filled three-band model (see the SM). We still adopt the bilayer two-orbital model in the following calculations.

Fig.\ref{fig2}(c) displays the Fermi surfaces with a filling of three electrons. The electron-like pocket $\alpha$ around the $\Gamma$ point is attributed to an almost equal mixture between $d_{x^2-y^2}$ and $d^+_{z^2}$ orbitals. Meanwhile, the two hole-like pockets $\beta$ and $\gamma$ are dominantly contributed by $d_{x^2-y^2}$ and $d^+_{z^2}$ orbitals, respectively. The large $d_{x^2-y^2}$ pocket $\beta$ is just the representative Fermi surface of cuprates, except for some moderate hybridization with the $d_{z^2}$ orbital along $\text{M}-\text{X}$. The pocket $\gamma$, originating from the inter-layer bonding state $d^+_{z^2}$, carries large density of states owing to the relative flat band dispersion around the M point. To characterize the Fermi surface nesting, we calculate the bare spin susceptibility and the largest eigenvalues of the susceptibility matrix $\chi_0(\bm{q})$ are shown in Fig.\ref{fig2}(d). The peak around the $\bm{q}_3=(\pi/2,\pi/2)$ vector is attributed to the Fermi surface nesting between electron-like $\alpha$ and hole-like $\gamma$ pockets. The peaks around the $\bm{q}_1=(0.47\pi,0)$ and $\bm{q}_2=(\pi,0)$ vectors are derived from the Fermi surface nesting between $\beta$ and $\gamma$ pockets, as shown in Fig.\ref{fig2} (d). Owing to the distinct shapes of the $\alpha$ ad $\gamma$ pockets, the nesting at $\bm{q}=(\pi,\pi)$ between them is not strong and this feature is even more prominent at $k_z=\pi$ plane in the DFT calculations (see Fig.\ref{fig1}(d) and SM).

\begin{figure}[t]
\centerline{\includegraphics[width=0.5\textwidth]{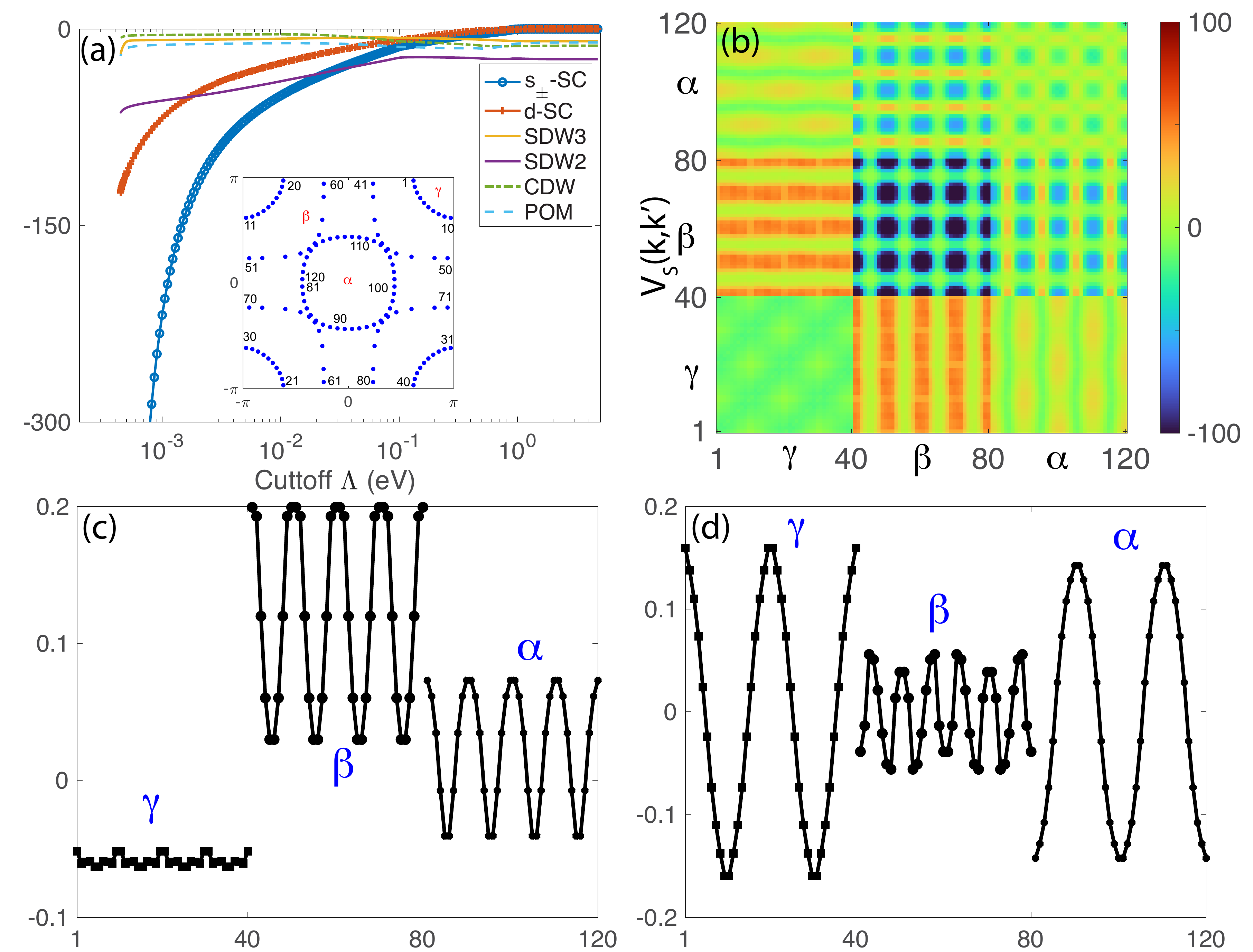}}
\caption{(color online). (a) Typical flow of leading instabilities in both particle-hole and particle-particle channels at $n=3.0$. SDW2 and SDW3 denote SDW states with the vectors of $\bm{q}_2$ and $\bm{q}_3$, respectively. The inset shows Fermi surface patching indices and the numbers 1-40, 41-80 and 81-120 represent patches on the $\gamma$, $\beta$ and $\alpha$ pockets, respectively. (b) Cooper pairing scattering interactions on Fermi surfaces at the critical $\Lambda_c$. Gap functions of the $s_{\pm}$-wave pairing (c) and $d_{x^2-y^2}$-wave pairing (d). The adopted parameters are $U$=5 eV and $J/U=0.1$ and the resulting maximum value of the initial vertex function in the band space (max$|\Gamma^{\Lambda_0}(\bm{k}_1,\bm{k}_2,\bm{k}_3)|$=1.25 eV) is still  located in the intermediate coupling regime.     \label{fig3}}
\end{figure}

{\it FRG analysis}. To investigate the pairing symmetry and other correlated electronic states in La$_3$Ni$_2$O$_7$, we first consider the weak-coupling to intermediate-coupling limit and adopt the FRG approach, which treats all particle-hole and particle-particle channels on equal footing~\cite{FWang2010,Metzner2012,Platt2013}. In our FRG calculations, we consider the on-site Hubbard intra- and inter-orbial repulsion, Hund's coupling, as well as pair-hopping interactions,
\begin{eqnarray}
H_{\text{int}}&=&U\sum_{i\alpha}n_{i\alpha\uparrow}n_{i\alpha\downarrow}
+U'\sum_{i,\alpha<\beta}n_{i\alpha}n_{i\beta}\nonumber\\
&+&J\sum_{i,\alpha<\beta,\sigma\sigma'}c^{\dag}_{i\alpha\sigma}c^{\dag}_{i\beta\sigma'}c_{i\alpha\sigma'}c_{i\beta\sigma}\nonumber\\
&+&J'\sum_{i,\alpha\neq\beta}c^{\dag}_{i\alpha\uparrow}c^{\dag}_{i\alpha\downarrow}c_{i\beta\downarrow}c_{i\beta\uparrow}
\label{interaction2}
\end{eqnarray}
where $n_{i\alpha}=n_{\alpha\uparrow}+n_{\alpha\downarrow}$. $U$, $U'$, $J$ and $J'$
represent the onsite intra- and inter-orbital repulsion and the onsite
Hund's coupling and pair-hopping terms for the Ni site, respectively. We use the Kanamori relations $U=U'+2J$ and $J=J'$. Within FRG formalism, we focus on the evolution of effective vertex function $\Gamma^\Lambda(\bm{k}_1,\bm{k}_2,\bm{k}_3)$ as a function of an energy scale by progressively integrating out high-energy modes. The vertex function can be decoupled into the summation of bilinear terms $\sum_{\xi \bm{k}_1\bm{k}_2} \Phi^{\Lambda,\xi}(\bm{k}_1,\bm{k}_2)\mathcal{O}^{\xi,\dagger}_{\bm{k}_1}(\bm{Q})\mathcal{O}^{\xi}_{\bm{k}_2}(\bm{Q})$, where $\mathcal{O}^{\xi}_{\bm{k}}(\bm{Q})$ is the favored order operator in the particle-hole or particle-particle channels and $\Phi^{\Lambda,\xi}$ is the corresponding $\xi$-channel coupling. This channel coupling be further decouled into $\Phi^{\Lambda,\xi}(\bm{k}_1,\bm{k}_2)=\sum_m \phi^\xi_m(\Lambda)f^\xi_m(\bm{k}_1)f^\xi_m(\bm{k}_2)$, where $\phi^\xi_m(\Lambda)$ is an  eigenvalue and $f^\xi_m(\bm{k})$ is the corresponding eigenmode, transforming as in irreducible representation of the symmetry group of $\phi^{\Lambda,\xi}(\bm{k}_1,\bm{k}_2)$. Approaching a critical energy scale $\Lambda_c$, $\Gamma^\Lambda(\bm{k}_1,\bm{k}_2,\bm{k}_3)$ tends to diverge and this signals an ordering tendency in the $\xi$ channel with the most divergent $\phi^\xi_m(\Lambda)$.

Fig.\ref{fig3}(a) displays the representative FRG flow with a decreasing energy scale $\Lambda$. Away from the Fermi level, the spin density wave (SDW) fluctuation with a vector $\bm{q}_2$ is established. With a decreasing cutoff $\Lambda$, the eigenvalues of two superconducting states grow rapidly, exceed the SDW state and finally diverge approaching to the Fermi level. This clearly suggests a spin-fluctuation-mediated pairing. At the critical $\Lambda_c$, the effective Cooper pairing scattering interaction $V_S(\bm{k},\bm{k}')$ in the spin-singlet channel is shown in Fig.\ref{fig3}(b). We find that the dominant inter-pocket Cooper pairing scattering occurs between the $\beta$ and $\gamma$ pockets and is repulsive, which predominantly determines the pairing symmetry. The dominant pairing state is $s_{\pm}$-wave and its gap function is shown in Fig.\ref{fig3}(c), where the gaps on the $\beta$ pocket are anisotropic and the gaps of the $\beta$ and $\gamma$ have the opposite signs. The gap of the $\alpha$ pocket exhibits a sign change owing to the sign-changed Cooper pairing scattering interaction between the $\alpha$ and $\beta$ pockets in the momentum space. As the $\gamma$ pocket is from the bonding state $d^+_{z^2}$ and the $\beta$ pocket is from the interlayer $d_{x^2-y^2}$ antibonding state, the sign reversal of the gap functions between these pockets resembles the $s_{\pm}$ pairing proposed in bilayer Hubbard model with a strong interlayer coupling~\cite{HZhai2009,TMaier2011,MNakata2017}. The subdominant pairing is $d_{x^2-y^2}$-wave with SC nodes along the diagonal direction, as shown in Fig.\ref{fig3}(d). The segments from the $\beta$ and $\gamma$ pockets connected by the nesting vector $\bm{q}_2$ still have the opposite superconducting gaps. With the increasing Coulomb repulsion, the system is in proximity to an SDW instability. We further study the external doping with a rigid band shift approximation and find the electron doping can enhance T$_c$ without changing the pairing symmetry.

\begin{figure}[t]
\centerline{\includegraphics[width=0.5\textwidth]{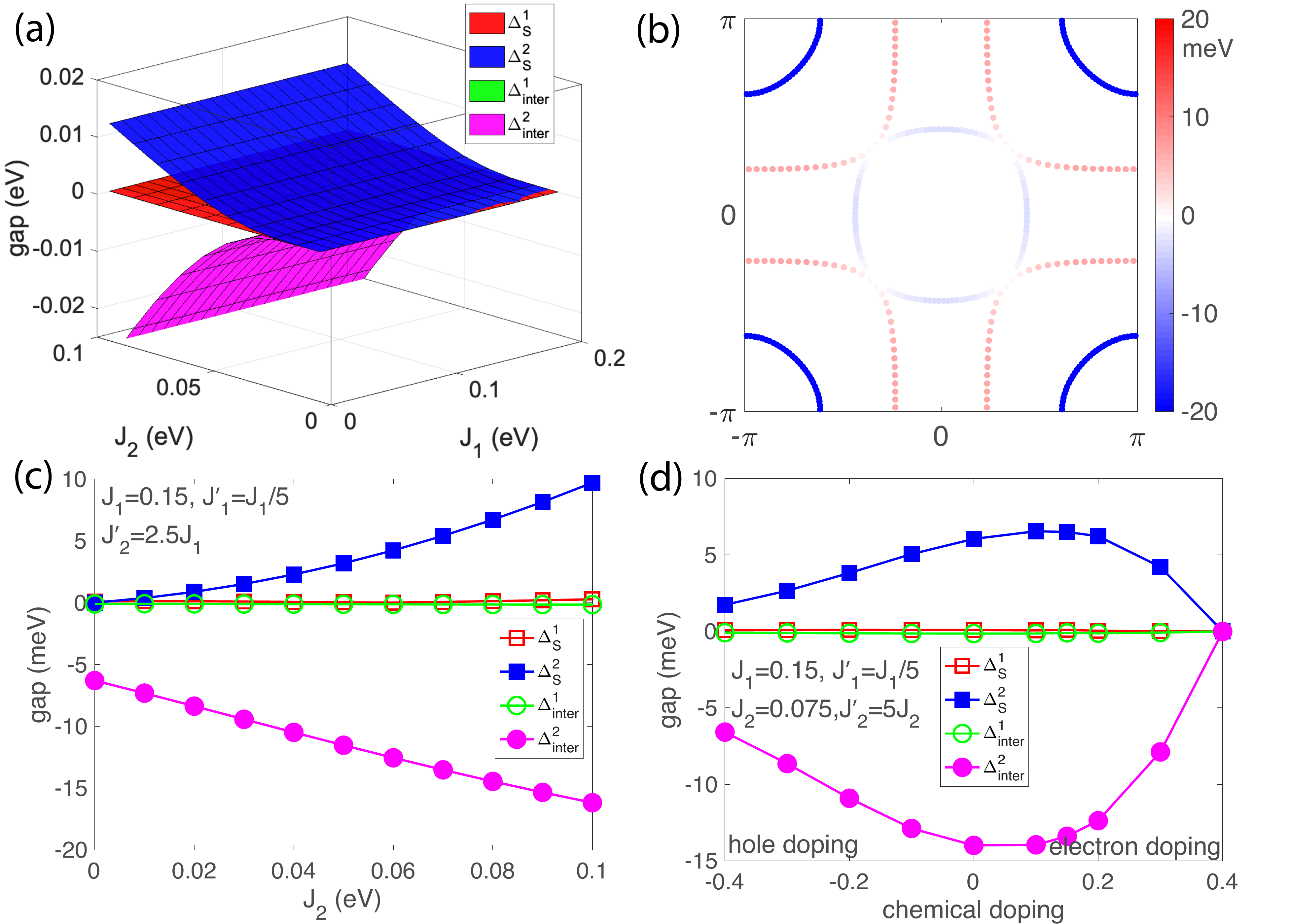}}
\caption{(color online). (a) Leading superconducting gaps in different channels as a function of exchange couplings $J_1$ and $J_2$. (b) Superconducting gaps on the Fermi surfaces with $J_1=0.15$ eV and $J_2=0.075$ eV. (c) Superconducting gaps as a function of $J_2$ with $J_1=0.15$ eV, $J'_1=J_1/5$ and $J'_2=2.5J_1$. (d) Superconducting gaps as a function of charge doping with the same parameters in (b). Interlayer exchange couplings $J'_1=J_1/5$ and $J'_2=5J_2$ are used in the calculations, unless explicitly specified. \label{fig4}}
\end{figure}

{\it Pairing in the t-J model}. In analogous to cuprates, we also investigate the pairing symmetry of La$_3$Ni$_2$O$_7$ from a strong-coupling perspective. We adopt the two-orbital t-J model to investigate pairing symmetries for nickelates similar to iron-based superconductors~\cite{SiQM2008,Seo2008} and consider both the in-plane and out-of-plane antiferromagnetic couplings between the spin of Ni $d_{x^2-y^2}/d_{z^2}$ orbitals,
\begin{eqnarray}
H_{J}=\sum_{\langle ij\rangle\alpha}J^\alpha_{ij}(\mathbf{S}_{i\alpha}\mathbf{S}_{j\alpha}-\frac{1}{4}n_{i\alpha}n_{j\alpha})
\end{eqnarray}
where
$\bm{S}_{i\alpha}=\frac{1}{2}c_{i\alpha\sigma}^{\dagger}\bm{\sigma}_{\sigma\sigma'}c_{i\alpha\sigma'}$
is the local spin operator and $n_{i\alpha}$is the local density
operator for Ni $\alpha$ orbital ($\alpha=1,2$). $\langle ij\rangle$ denotes the in-plane and out-of-plane nearest neighbors and the in-plane and out-of-plane couplings are $J^{\alpha}_{x/y}=J_\alpha$ and $J^{\alpha}_{z}=J'_\alpha$, respectively.   By performing the Fourier transformation, $H_{J}$ in momentum space reads
\begin{eqnarray}
H_{J}&=&\sum_{\eta\alpha\mathbf{k}\bm{k}'}V^\alpha_{\mathbf{k},\mathbf{k}'}P^\dag_{\eta\alpha }(\bm{k})P_{\eta\alpha}(\bm{k}')
+\sum_{\bm{k}\bm{k}'}W^{\alpha}_{\bm{k},\bm{k}'}B^\dag_{\alpha}(\bm{k})B_{\alpha}(\bm{k}'),\nonumber\\
\end{eqnarray}
with the intralayer pair operator $P^\dag_{\eta\alpha }(\bm{k})=c_{\eta\alpha\uparrow}^{\dagger}(\bm{k})c_{\eta\alpha\downarrow}^{\dagger}(-\bm{k})$ and the interlayer pair operator $B^\dag_{\alpha}(\bm{k})=c_{t\alpha\uparrow}^{\dagger}(\bm{k})c_{b\alpha\downarrow}^{\dagger}(-\bm{k})+c_{b\alpha\uparrow}^{\dagger}(\bm{k})c_{t\alpha\downarrow}^{\dagger}(-\bm{k})$. Here $V^\alpha_{\mathbf{k},\mathbf{k}'}=-\frac{2J_\alpha}{N}\sum_{\pm}(cosk_x\pm cosk_y)(cosk'_x\pm cosk'_y)$ and $W^\alpha_{\mathbf{k},\mathbf{k}'}=-\frac{J'_\alpha}{2N}$ and $N$ being the number of lattice sites. We investigate the pairing state for both undoped and doped systems and neglect the
no-double-occupancy constraint on this t-J model and then perform a mean-field decoupling and solve the self-consistent gap equations (details in SM). According to our calculations, we find that the pairing state has weak dependence on the interlayer and intralayer exchange couplings for the $d_{x^2-y^2}$ orbital. By choosing $J'_1/J_1=1/5$ and $J'_2/J_2=5$ due to the strong interlayer coupling in the $d_{z^2}$ orbitals, the obtained superconducting gaps in different channels as a function of $J_1$ and $J_2$ are shown in Fig.\ref{fig4}(a). We find that $J'_2$ promotes a dominant interlayer pairing and the $J_2$ generates a substantial extended $s$-wave intralayer pairing in the $d_{z^2}$ orbital. While, the pairing in the $d_{x^2-y^2}$ orbital remains relatively weak. Intriguingly, the interlayer pairing in the $d_{z^2}$ orbital consistently exhibits an opposite sign to its intralayer counterpart. With increasing exchange couplings, the $s$-wave superconducting gaps and the associated condensation energy both grows rapidly, while the $d$-wave pairing is not favored. By choosing $J_1=0.15$ eV and $J_2=0.075$ eV, the obtained superconducting gaps are displayed in Fig.\ref{fig4}(b). It is apparent that the superconducting gap on the $\gamma$ pocket dominates over other pockets and $\alpha$, $\gamma$ and $\beta$ pockets feature sign-reversal superconducting gaps, in analogous to the $s_{\pm}$-wave pairing in iron based superconductors~\cite{iron-based_2008,RMP.83.1589} and the bilayer Hubbard model~\cite{HZhai2009,TMaier2011,MNakata2017}. Quasi nodal features appear along the diagonal direction on both $\alpha$ and $\beta$ pockets. Since the extended $s$-wave form factor is negative around the M point, the opposite signs of the intralayer pairing $\Delta^2_S$ and interlayer pairing $\Delta^2_{\text{inter}}$ in the $d_{z^2}$ orbital can lead to an enhancement of the gap on the $\gamma$ pocket, which has a large density of states, making this configuration favorable. We further explore the effect of the in-plane exchange coupling $J_2$ on the pairing by varying $J_2$ while keeping other parameters fixed. As shown in Fig.\ref{fig4}(c), the superconducting gaps are greatly reduced by setting $J_2=0$. An increasing $J_2$ not only promotes the intralayer pairing for the $d_{z^2}$ orbital but also significantly enhances the interlayer pairing $\Delta^2_{\text{inter}}$. A strong $J_2$ will generate a positive gap on the $\alpha$ pocket and the superconducting gaps on the $\alpha$ and $\gamma$ have the opposite sign, similar to that proposed in CuO$_2$ monolayer~\cite{KJiang2018}. Moreover, we study the effect of external doping on the  gap amplitude and the obtained gap evolution as a function of doping is depicted in Fig.\ref{fig4}(d) with $J_1=0.15$ eV and $J_2=0.075$ eV. The optimal doping is achieved around the electron doping $\delta=0.1$, where the gap amplitudes reach the maximum. The $s_{\pm}$-wave pairing vanishes when the $\gamma$ pocket is absent around the heavy electron doping $\delta=0.4$. The obtained pairing state and doping dependent gap are qualitatively consistent with our FRG results despite some quantitative difference in gap distribution on Fermi surfaces.

{\it Discussion}. In our calculations, we have neglected the orthorhombic distortion in the high-pressure phase. As the orthorhombic distortion is along the diagonal direction, the induced mixture of $s$- and $d$-wave pairing between the NN sites is expected to be not strong. The $s_{\pm}$-wave pairing is still dominant according to our calculations. In DFT calculations, the $d_{z^2}$ bands exhibit noticeable dispersion and the $\gamma$ pocket is elongated along the diagonal direction with increasing $k_z$, resulting a flower-shape pocket at the $k_z=\pi$ plane (see Fig.\ref{fig1}(d) and SM). To deal with the 3D feature, we also construct a 3D TB model and perform calculations at different $k_z$ planes with the t-J model. We find that the $s_{\pm}$-wave pairing is always dominant. Both our FRG and t-J model calculations show that electron doping can enhance T$_c$. This observation aligns with the experimental observation that superconducting samples often exhibit oxygen deficiency~\cite{sun2023}, leading to electron doping. As the $s_{\pm}$-wave state possesses sign-reversal superconducting gaps in momentum space, neutron scattering and scanning tunneling microscopy measurements are helpful to identify the pairing symmetry~\cite{Hirschfeld_2011,RN3410,Hanaguri2010}.  

Our study highlights the pivotal role of the $d_{z^2}$ orbital in promoting superconductivity in La$_3$Ni$_2$O$_7$.  The $\gamma$ pocket from the $d_{z^2}$ bonding states is crucial for mediating superconductivity and the strong $d_{z^2}$ interlayer pairing originates from the strong hybridization between the Ni-$d_{z^2}$ orbital and the apical O-$p_z$ orbital. A finite intralayer exchange coupling $J_2$ can dramatically enhance both intralayer and interlayer the superconducting gaps. These tend to suggest that both interlayer and intralayer exchange couplings of the $d_{z^2}$ orbital are important for the observed high-T$_c$ superconductivity in La$_3$Ni$_2$O$_7$, distinct from the bilayer cuprates. These may provide an explanation for the observed high-T$_c$ superconductivity in La$_3$Ni$_2$O$_7$ with pressure. Superconductivity is only observed at a high pressure where the metallization of the bonding $d_{z^2}$ state is achieved~\cite{sun2023} and the $\gamma$ pocket emerges accordingly. With an increasing pressure, the in-plane couplings for both $d_{x^2-y^2}$ and $d_{z^2}$ orbitals get strengthened. Given that the intralayer exchange coupling is strong for the $d_{x^2-y^2}$ orbital but weak for the $d_{z^2}$ orbital, the external pressure will noticeably enhance the intralayer exchange coupling for the $d_{z^2}$ orbital, which dramatically promotes superconductivity according to our calculations.

{\it Conclusion}. In conclusion, we study the low-energy electronic structures of the recently discovered high T$_c$ superconductor La$_3$Ni$_2$O$_7$ and study the pairing symmetry based on a bilayer two-orbital model from both weak-coupling and strong-coupling perspective. Under the standard methods, an $s_{\pm}$-wave pairing with sign-reversal gaps on different Fermi surfaces is identified. Our work underscores the pivotal role of the Ni-$d_{z^2}$ orbital in driving superconducting pairing, with both interlayer and intralayer exchange couplings being essential for the attainment of high-T$_c$ superconductivity in La$_3$Ni$_2$O$_7$. Our study provides valuable insights into key ingredients for high T$_c$ superconductivity in La$_3$Ni$_2$O$_7$ with pressure and contributes to the understanding
of unconventional superconductors.

{\it Acknowledgments}. We acknowledge the supports by the Ministry of Science and Technology (Grant No. 2022YFA1403901), National Natural Science Foundation of China (No. 11920101005, No. 11888101, No. 12047503) and the New Cornerstone Investigator Program. Yuhao Gu also acknowledges the supports from China Postdoctoral Science Foundation Fellowship (No.2023T160675).

{\it Note added}. During the preparation of this work, we became aware of several independent
studies of electronic structure and pairing instabilities about La$_3$Ni$_2$O$_7$~\cite{zhang2023electronic,yang2023possible,Frank2023electronic,Hirofumi2023possible}. Our pairing state is consistent with Ref.\cite{yang2023possible} and the T$_c$ evolution with doping is consistent with Ref.\cite{Hirofumi2023possible}.

\bibliography{references_new0831}
 \bibliographystyle{apsrev4-1}

\appendix
\clearpage

\section{Computational methods}
Our electronic structure calculations employ the Vienna ab initio simulation package (VASP) code\cite{kresse1996} with the projector augmented wave (PAW) method\cite{Joubert1999}. The Perdew-Burke-Ernzerhof (PBE)\cite{perdew1996} exchange-correlation functional is used in our calculations. The kinetic energy cutoff is set to be 600 eV for the expanding the wave functions into a plane-wave basis and the energy convergence criterion is $10^{-7}$ eV. We employ the primitive cells of \ce{La3Ni2O7} under under ambient pressure and high pressure to perform calculations. We optimize the atomic positions until the force is less than 0.005 eV/\AA with the experimental lattice constants. The $\Gamma$-centered \textbf{k}-meshes are $9\times9\times9$ and $14\times14\times14$ and for ambient pressure phase and high pressure phase, respectively.

We employ Wannier90\cite{mostofi2008wannier90,Marzari2012} to calculate maximally localized Wannier functions and the disentangled bands. The initial projectors are Ni's $d_{x^2-y^2}+d_{z^2}$ orbitals. We fit the effective tight-binding model with the data of the disentangled bands near the Fermi energy.

\section{electronic structures under ambient pressure and high pressure}

\begin{figure}
\centerline{\includegraphics[width=0.5\textwidth]{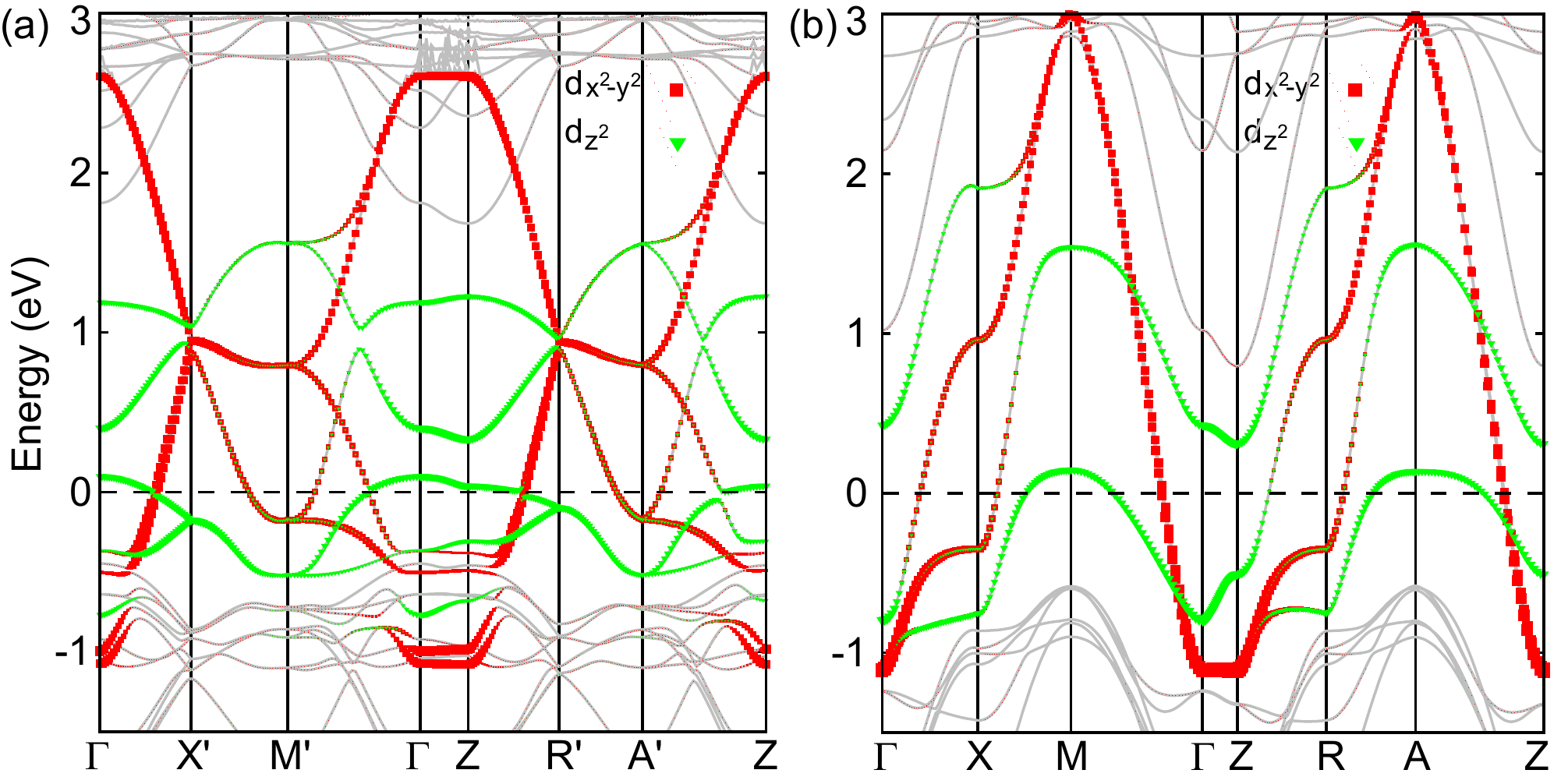}}
\caption{(color online) The DFT band structures of \ce{La3Ni2O7} in the primitive cell under (a) ambient pressure and (b) high pressure. The orbital characters near fermi level are represented by different colors. \label{SM1}}
\end{figure}

In the compound \ce{La3Ni2O7}, the phenomenon of high-temperature superconductivity has been experimentally observed to manifest specifically within its high-pressure phase, as opposed to its ambient-pressure phase. In order to elucidate the distinctions between these phases, Fig.\ref{SM1} (a) and (b) illustrate band structures under both ambient pressure and high pressure. On one hand, these two phases clearly reveal the presence of the bonding state $d_{z^2}^+$ band and the anti-bonding state $d_{z^2}^-$ band. On the other hand, due to the further suppression of octahedral tilting within the high-pressure phase\cite{sun2023}, the bands originating from the $d_{x^2-y^2}$ orbital, the bonding and anti-bonding states $d_{z^2}^{\pm}$, exhibit an increased band width.  Additionally, a slight enhancement is observed in the energy splitting between the bonding and anti-bonding state bands.

\begin{figure}
\centerline{\includegraphics[width=0.5\textwidth]{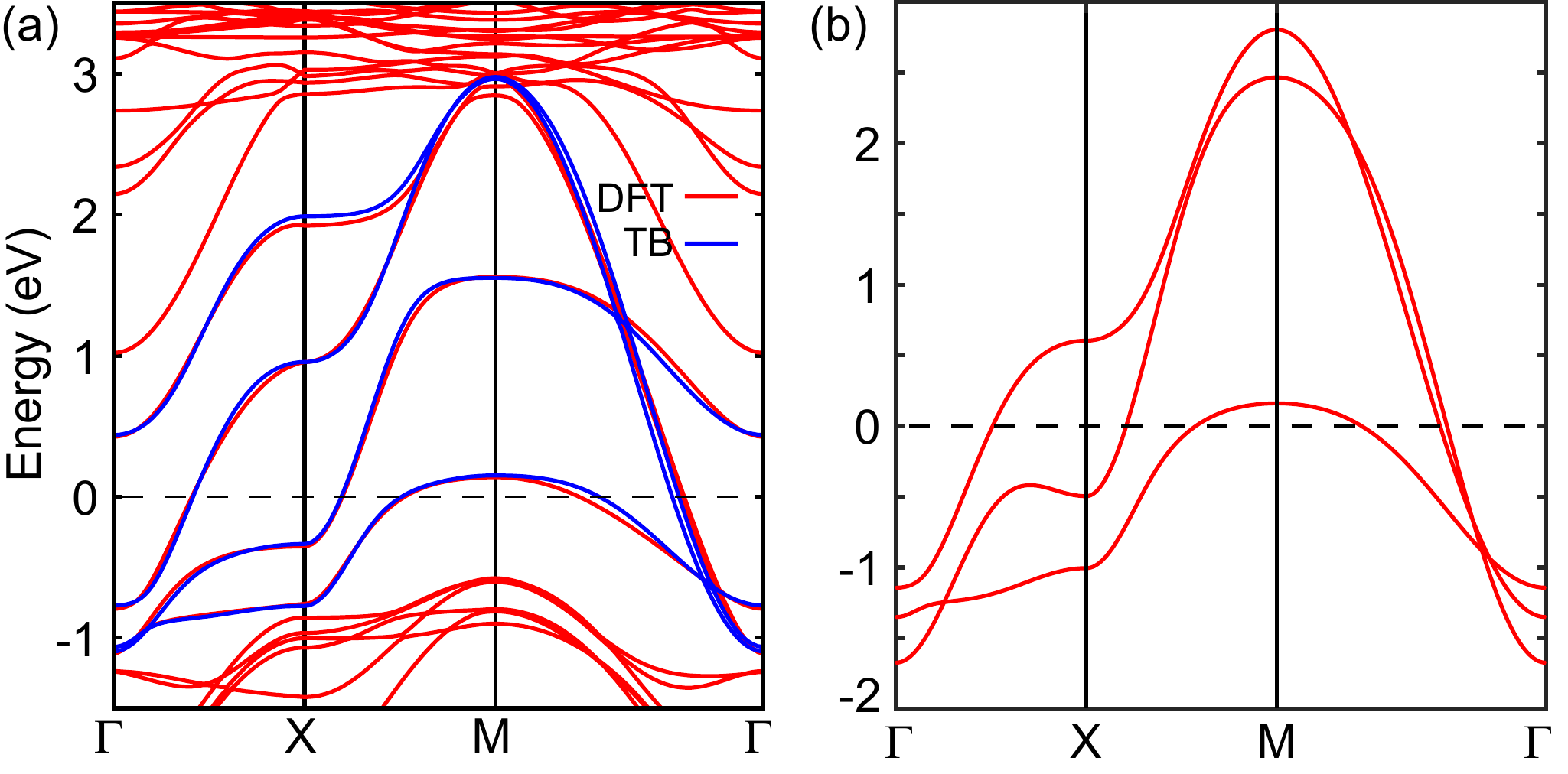}}
\caption{(color online). (a) The band structures from DFT calculations and bilayer two-band model. Noting that the chemical potential of TB result is shifted to reproduce the DFT result. (b) The band structure of three-band effective model.  \label{SM2}}
\end{figure}

\section{bilayer two-orbital model and Three band model on the square lattice}

The matrix representation of the bilayer two-band model discussed on the square lattice in the main text can be expressed as follows:

\begin{eqnarray}
h(\mathbf{k}) =\left(\begin{array}{cccc}
h_{11}(\mathbf{k}) & h_{12}(\mathbf{k}) & h_{13}(\mathbf{k}) & h_{14}(\mathbf{k})   \\
h_{12}^*(\mathbf{k}) & h_{22}(\mathbf{k})  & h_{23}(\mathbf{k}) & h_{24}(\mathbf{k}) \\
h_{13}^*(\mathbf{k}) & h_{23}^*(\mathbf{k}) &h_{33}(\mathbf{k}) & h_{34}(\mathbf{k})  \\
h_{14}^*(\mathbf{k}) &  h_{24}^*(\mathbf{k}) & h_{34}^*(\mathbf{k}) & h_{44}(\mathbf{k})   \\
\end{array}\right),
\end{eqnarray}

where the basis operator is denoted by $\psi_{\mathbf{k} \sigma}^{\dagger}=[c_{t x^2 \sigma}^{\dagger}(\mathbf{k}), c_{t z^2 \sigma}^{\dagger}(\mathbf{k}), c_{b x^2 \sigma}^{\dagger}(\mathbf{k}), c_{b z^2 \sigma}^{\dagger}(\mathbf{k})]$. The band structures obtained from density functional theory (DFT) calculations and the bilayer two-band model are presented in Fig.\ref{SM2}(a), revealing a close agreement near the Fermi level. Then, the matrix elements in the Hamiltonian $h(\mathbf{k})$ matrix can be given by

\begin{eqnarray}
h_{11}(\boldsymbol{k})&=&h_{33}(\boldsymbol{k})=\epsilon_{x^2}+2 t_{11}^x\left(\cos k_x+\cos k_y\right)\nonumber
\\
&+&4 t_{11}^{x y} \cos k_x \cos k_y+2 t_{11}^{x x}\left(\cos 2 k_x+\cos 2 k_y\right),\nonumber
\\
h_{22}(\boldsymbol{k})&=&h_{44}(\boldsymbol{k})=\epsilon_{z^2}+2 t_{22}^x\left(\cos k_x+\cos k_y\right)\nonumber
\\
&+&4 t_{22}^{x y} \cos k_x \cos k_y+2 t_{22}^{x x}\left(\cos 2 k_x+\cos 2 k_y\right),\nonumber
\\
h_{12}(\boldsymbol{k})&=&2 t_{12}^x\left(\cos k_x-\cos k_y\right)+2 t_{12}^{x x}\left(\cos 2 k_x-\cos 2 k_y\right),\nonumber
\\
h_{13}(\boldsymbol{k})&=&s_{11}^0+2 s_{11}^x\left(\cos k_x+\cos k_y\right)+4 s_{11}^{x y} \cos k_x \cos k_y
\nonumber
\\
&+&2 s_{11}^{x x}\left(\cos 2 k_x+\cos 2 k_y\right),\nonumber
\\
h_{14}(\boldsymbol{k})&=&2 s_{12}^x\left(\cos k_x-\cos k_y\right)+2 s_{12}^{x x}\left(\cos 2 k_x-\cos 2 k_y\right),\nonumber
\\
h_{24}(\boldsymbol{k})&=&s_{22}^0+2 s_{22}^x\left(\cos k_x+\cos k_y\right)+4 s_{22}^{x y} \cos k_x \cos k_y\nonumber
\\
&+&2 s_{22}^{x x}\left(\cos 2 k_x+\cos 2 k_y\right),\nonumber
\\
h_{23}(\boldsymbol{k})&=&h_{14}(\boldsymbol{k}),h_{34}(\boldsymbol{k})=h_{12}(\boldsymbol{k})
\end{eqnarray}

The corresponding tight binding parameters are specified in unit of eV as

\begin{eqnarray}
&& \epsilon_{x^2}=10.929, t_{11}^x=-0.505,t_{11}^{x y}=0.060,t_{11}^{x x}=-0.047,\nonumber  \\
&& \epsilon_{z^2}=10.474,t_{22}^x=-0.126,t_{22}^{x y}=-0.003,t_{22}^{x x}=-0.021,\nonumber   \\
&& s_{11}^0=-0.049,s_{11}^x=0.001,s_{11}^{x y}=0.035,s_{11}^{x x}=-0.022,\nonumber  \\
&& s_{22}^0=-0.628,s_{22}^x=0.011, s_{22}^{x y}=-0.032, s_{12}^{x x}=0.024,\nonumber  \\
&& t_{12}^x=0.253,t_{12}^{x x}=0.037,s_{12}^x=-0.038,s_{12}^{x x}=-0.009,
\end{eqnarray}

where $t^m_{\alpha\beta}$ and $s^m_{\alpha\beta}$ (m = 1, 2) represent the intralayer and interlayer hopping, respectively. According to the valence of Ni$^{2.5+}$, the electron filling in our model is $n=3$ with $\mu = 10.0585 eV$. Since the band structures at the fermi level are predominantly attributed to d$_{x^2-y^2}$ orbital and bonding state d$^{+}_{z^2}$, a unitary transformation can be applied to the aforementioned effective model to provide a more accurate description of the low-energy physics. As a result, the effective model is transformed into the following form:

\begin{widetext}
\begin{eqnarray}
&&h^{\prime}(\mathbf{k})=Uh(\mathbf{k})U^{-1}\nonumber
\\
&&=\left(\begin{array}{cccc}
h_{11} & \frac{1}{\sqrt{2}}[h_{12}+h_{14}] & h_{13} & \frac{1}{\sqrt{2}}[h_{12}-h_{14}]   \\
\frac{1}{\sqrt{2}}[h^*_{12}+h^*_{14}] & \frac{1}{2}[h_{22}+h_{24}+h^*_{24}+h_{44}]  & \frac{1}{\sqrt{2}}[h_{23}+h^*_{34}] & 0 \\
h_{13}^* & \frac{1}{\sqrt{2}}[h^*_{23}+h_{34}] &h_{33} & \frac{1}{\sqrt{2}}[h^*_{23}-h_{34}]  \\
\frac{1}{\sqrt{2}}[h^*_{12}-h^*_{14}] & 0 & \frac{1}{\sqrt{2}}[h_{23}-h^*_{34}] &  \frac{1}{2}[h_{22}-h_{24}-h^*_{24}+h_{44}]   \\
\end{array}\right),U =\left(\begin{array}{cccc}
1 & 0 & 0 & 0   \\
0 & \frac{1}{\sqrt{2}}  & 0 & \frac{1}{\sqrt{2}} \\
0 & 0 &1 & 0  \\
0 & \frac{1}{\sqrt{2}} & 0 & -\frac{1}{\sqrt{2}}   \\
\end{array}\right).
\end{eqnarray}
\end{widetext}
Then, the corresponding basis operator is transformed to $\psi_{\mathbf{k} \sigma}^{\dagger\prime}=\left[c_{t x^2 \sigma}^{\dagger}(\mathbf{k}), c_{+,z^2 \sigma}^{\dagger}(\mathbf{k}), c_{b x^2 \sigma}^{\dagger}(\mathbf{k}), c_{-,z^2 \sigma}^{\dagger}(\mathbf{k})\right]$ with $c_{\pm,z^2 \sigma}^{\dagger}(\mathbf{k})=\frac{1}{\sqrt{2}}(c_{t z^2 \sigma}^{\dagger}(\mathbf{k})\pm c_{b z^2 \sigma}^{\dagger}(\mathbf{k}))$. Considering that the band structure associated with the antibonding state d$^{-}_{z^2}$ is located away from the Fermi level, it can be neglected, leading to the further simplification of the bilayer two-orbital model into a three-band model. The three-band model can be written as $\tilde{\mathcal{H}}_{\mathrm{TB}}=\sum_{\mathbf{k} \sigma} \tilde{\psi}_{\mathbf{k} \sigma}^{\dagger}[\tilde{h}(\mathbf{k})-\tilde{\mu}] \tilde{\psi}_{\mathbf{k} \sigma}$, where the matrix elements in the Hamiltonian $\tilde{h}(\mathbf{k})$ matrix are given by

\begin{eqnarray}
\tilde{h}_{11}(\boldsymbol{k})&=&\tilde{h}_{33}(\boldsymbol{k})=\tilde{\epsilon}_{x^2}+2 \tilde{t}_{11}^x\left(\cos k_x+\cos k_y\right)\nonumber
\\
&+&4 \tilde{t}_{11}^{x y} \cos k_x \cos k_y+2 \tilde{t}_{11}^{x x}\left(\cos 2 k_x+\cos 2 k_y\right),\nonumber
\\
\tilde{h}_{22}(\boldsymbol{k})&=&\tilde{\epsilon}_{z^2}+2 \tilde{t}_{22}^x\left(\cos k_x+\cos k_y\right)\nonumber
\\
&+&4 \tilde{t}_{22}^{x y} \cos k_x \cos k_y+2 \tilde{t}_{22}^{x x}\left(\cos 2 k_x+\cos 2 k_y\right),\nonumber
\\
\tilde{h}_{12}(\boldsymbol{k})&=&2 \tilde{t}_{12}^x\left(\cos k_x-\cos k_y\right)+2 \tilde{t}_{12}^{x x}\left(\cos 2 k_x-\cos 2 k_y\right),\nonumber
\\
\tilde{h}_{13}(\boldsymbol{k})&=&\tilde{s}_{11}^0+2 \tilde{s}_{11}^x\left(\cos k_x+\cos k_y\right)+4 \tilde{s}_{11}^{x y} \cos k_x \cos k_y
\nonumber
\\
&+&2 \tilde{s}_{11}^{x x}\left(\cos 2 k_x+\cos 2 k_y\right),\nonumber
\\
\tilde{h}_{23}(\boldsymbol{k})&=&\tilde{h}_{12}(\boldsymbol{k})
\end{eqnarray}

The corresponding tight binding parameters are specified in unit of eV as

\begin{eqnarray}
&& \tilde{\epsilon}_{x^2}=10.746,~ \tilde{t}_{11}^x=-0.518,~\tilde{t}_{11}^{x y}=0.097,~ \tilde{t}_{11}^{x x}=-0.079,\nonumber  \\
&& \tilde{\epsilon}_{z^2}=9.869,~ \tilde{t}_{22}^x=-0.163,~ \tilde{t}_{22}^{x y}=-0.004,~ \tilde{t}_{22}^{x x}=-0.022,\nonumber   \\
&& \tilde{s}_{11}^0=0.05,~ \tilde{s}_{11}^x=-0.001,~ \tilde{s}_{11}^{x y}=-0.014,~ \tilde{s}_{11}^{x x}=0.043,\nonumber  \\
&& \tilde{t}_{12}^x=0.134,~ \tilde{t}_{12}^{x x}=0.034,
\end{eqnarray}

Based on the valence of Ni$^{2.5+}$, the electron filling in our model corresponds to $n=3$, with $\tilde{\mu} = 10.256 eV$. The band structure is depicted in Fig.\ref{SM2} (b), and it demonstrates good agreement with the results obtained from DFT calculations.

\section{Tight-binding model with $k_z$ dispersion}
\begin{figure}
\centerline{\includegraphics[width=0.5\textwidth]{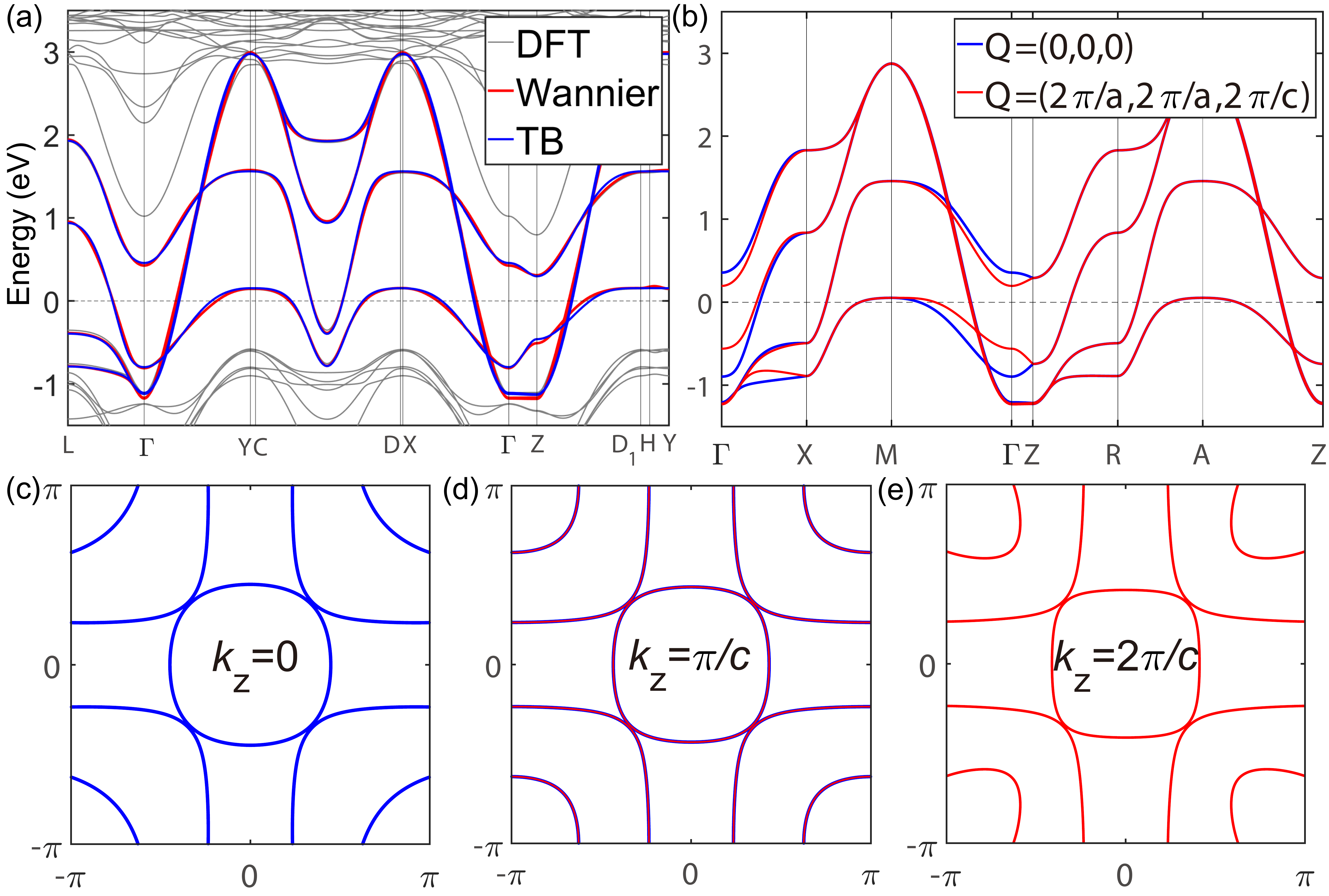}}
\caption{(color online) (a) The band structures of $Fmmm$ phase \ce{La3Ni2O7} from DFT calculation (gray lines), disentangled Wannierization (red lines) and the $k_z$-dependent TB model (blue lines). Noting that the chemical potentials of Wannierization and TB results are shifted to reproduce the DFT result. The notations of high symmetry $k$-points for $Fmmm$ phase \ce{La3Ni2O7} are same as the notations of reference's\cite{setyawan2010high}. (b) The band structure of  the block-diagonalized $k_z$-dependent TB model. The high symmetry $k$-points here are set as in an approximate tetragonal unit cell with 4 Ni atoms (neglecting the orthorhombic distortion of $Fmmm$ phase \ce{La3Ni2O7}). The chemical potential $\mu$ is 10.1277 eV here as a result of $n=3$ electron filling. The blue/red lines are the folded energy bands with a folding vector $Q=0/Q=(2\pi/a,2\pi/a,2\pi/c)$. (c-e) The Fermi surfaces of the block-diagonalized $k_z$-dependent TB model in (c) $k_z=0$ plane, (d) $k_z=\pi/c$ plane and (e) $k_z=2\pi/c$ plane. The blue/red lines represent the folded Fermi surfaces with a folding vector $Q=0/ Q=(2\pi/a,2\pi/a,2\pi/c)$.
\label{S3}}
\end{figure}

In order to describe $k_z$ dispersion, we have to consider inter-bilayer hopping in \ce{La3Ni2O7}. If we neglect the orthorhombic distortion of $Fmmm$ phase \ce{La3Ni2O7}, Ni atoms form the square lattice in \ce{NiO2} plane. As a result, there are four \ce{NiO2} layers located at $z=0,\delta_z,\frac{1}{2},\frac{1}{2}+\delta_z$ in one approximate tetragonal conventional unit cell, whose lattice constant is $\Vec{a_1}=(a,0,0), \Vec{a_2}=(0,a,0), \Vec{a_3}=(0,0,c)$ ($a$ is defined as the distance between two NN Ni atoms in \ce{NiO2} plane in this section). Then, the the basis operator is extended to 
\begin{eqnarray}
\psi_{\mathbf{k} \sigma}^{\dagger}&=&[c_{t x^2 \sigma}^{\dagger}(\mathbf{k}), c_{t z^2 \sigma}^{\dagger}(\mathbf{k}), c_{b x^2 \sigma}^{\dagger}(\mathbf{k}), c_{b z^2 \sigma}^{\dagger}(\mathbf{k}), \nonumber \\
& &c_{t' x^2 \sigma}^{\dagger}(\mathbf{k}), c_{t' z^2 \sigma}^{\dagger}(\mathbf{k}), c_{b' x^2 \sigma}^{\dagger}(\mathbf{k}), c_{b' z^ \sigma}^{\dagger}(\mathbf{k})].\\ \nonumber
\end{eqnarray}.

The $k_z$-dependent TB model has such form:
\begin{equation}
H_{3D}(\textbf{k})=
    \begin{pmatrix}
        H_{2D} & H_{z} \\
        H_{z}^{\dagger} & H_{2D}\\
    \end{pmatrix},
\end{equation}
where $H_{2D}$ is the intra-bilayer part. $H_{2D}$ is similar to our four-band TB model in the main text except adding the $k_z$-dependent term to the inter-layer hopping matrix elements:
\begin{eqnarray}
H_{2D,13}(\boldsymbol{k})&=&h_{13}(\boldsymbol{k})e^{i\delta_zk_z},\nonumber \\
H_{2D,14}(\boldsymbol{k})&=&h_{14}(\boldsymbol{k})e^{i\delta_zk_z},\nonumber \\
H_{2D,24}(\boldsymbol{k})&=&h_{24}(\boldsymbol{k})e^{i\delta_zk_z},\nonumber \\
H_{2D,23}(\boldsymbol{k})&=&H_{2D,14}(\boldsymbol{k}),\nonumber \\
\end{eqnarray}
and other matrix elements are identical to the corresponding ones in $h(\boldsymbol{k})$.

$H_z$ includes the hopping between two \ce{Ni2O4} bilayers:
\begin{eqnarray}
H_{z}(\mathbf{k}) =\left(\begin{array}{cccc}
H_{11}(\mathbf{k})  & 0                    & H_{13}(\mathbf{k})& 0   \\
0                   & H_{22}(\mathbf{k})   & 0                 & H_{24}(\mathbf{k}) \\
H_{13}^*(\mathbf{k})& 0                    & H_{11}(\mathbf{k})& 0  \\
0                   & H^*_{24}(\mathbf{k}) & 0                 & H_{22}(\mathbf{k})\\
\end{array}\right),
\end{eqnarray}
whose matrix elements are given by:
\begin{eqnarray}
H_{z,11}(\boldsymbol{k})&=8s^{z}_{11}cos(\frac{k_z}{2})cos(\frac{k_x}{2})cos(\frac{k_y}{2}),\nonumber \\
H_{z,22}(\boldsymbol{k})&=8s^{z}_{22}cos(\frac{k_z}{2})cos(\frac{k_x}{2})cos(\frac{k_y}{2}),\nonumber \\
H_{z,13}(\boldsymbol{k})&=4s^{z}_{13}e^{-i(1/2-\delta_z)k_z}cos(\frac{k_x}{2})cos(\frac{k_y}{2}),\nonumber \\
H_{z,24}(\boldsymbol{k})&=4s^{z}_{24}e^{-i(1/2-\delta_z)k_z}cos(\frac{k_x}{2})cos(\frac{k_y}{2}).\nonumber \\
\end{eqnarray}
Here $s^z_{11/22}$ describes the intra-orbital hopping between layer $z=0$ and layer $z=\frac{1}{2}$ and $s^z_{13/24}$ describes the intra-orbital hopping between layer $z=0$ and layer $z=\frac{1}{2}+\delta_z$.

As shown in Fig.\ref{S3}, we fit the TB parameters with Wannierization bands in AFLOW's $k$-path\cite{setyawan2010high} and they are specified in unit of eV as:
\begin{eqnarray}
&& \epsilon_{x^2}=10.920, t_{11}^x=-0.512,t_{11}^{x y}=0.062,t_{11}^{x x}=-0.053,\nonumber  \\
&& \epsilon_{z^2}=10.501,t_{22}^x=-0.123,t_{22}^{x y}=-0.005,t_{22}^{x x}=-0.022,\nonumber   \\
&& s_{11}^0=-0.030,s_{11}^x=0.001,s_{11}^{x y}=0.023,s_{11}^{x x}=-0.015,\nonumber  \\
&& s_{22}^0=-0.601,s_{22}^x=0.025, s_{22}^{x y}=-0.011, s_{12}^{x x}=0.011,\nonumber  \\
&& t_{12}^x=0.250,t_{12}^{x x}=0.035,s_{12}^x=-0.038,s_{12}^{x x}=-0.006, \nonumber \\ 
&& s^{z}_{11}=0.001,s^{z}_{13}=0.0004,s^{z}_{22}=-0.006,s^{z}_{24}=-0.031. \nonumber\\
\end{eqnarray}
According to the valence of \ce{Ni^{2.5+}}, the electron filling here is $n=3$ with $\mu=10.1277$ eV. We note that the value of the chemical potential is near to that in our DFT calculation ($\mu_{DFT}=10.1367$ eV).

We note that we can connect two bilayers with one translational symmetry $\frac{\Vec{a_1}}{2}+\frac{\Vec{a_2}}{2}+\frac{\Vec{a_3}}{2}$ in the approximate conventional tetragonal unit cell. As a result, we can transfer the $8\times8$ $k_z$-dependent TB model into a block-diagonalized matrix with using the translational symmetry:
\begin{equation}
H^{eff}_{3D}(\textbf{k})=
    \begin{pmatrix}
        H_k & 0 \\
        0 & H_{k+Q}\\
    \end{pmatrix},
\end{equation}
here $H_k$ is the effective four-band model and $Q=(2\pi/a,2\pi/a,2\pi/c)$. The explicit forms of $H_k$ and $H_{k+Q}$ are
\begin{eqnarray}
H_k = H_{2D}+H_{z}, H_{k+Q} = H_{2D}-H_{z}.
\end{eqnarray}
We plot the energy dispersion of each part in Fig.\ref{S3}(b), which reproduce the energy dispersion of our 8-band model. We also plot the Fermi surfaces on $k_z=0$ plane, $k_z=\pi/c$ plane and $k_z=2\pi/c$ plane, as shown in Fig.\ref{S3}(d-f). As our 8-band TB model describes the conventional cell, the Fermi surfaces/bands on $k_z=\pi/c$ plane folds to $k_z=-\pi/c$ plane, making the Fermi surfaces/bands doubly degenerate on $k_z=\pm\pi/c$ plane, as shown in Fig.\ref{S3}(b/d).

\section{pairing from the t-J model}

In the strong-coupling limit, we adopt the two-orbital t-J model to investigate pairing symmetries for nickelates similar to iron-based superconductors~\cite{SiQM2008,Seo2008} and consider both the in-plane and out-of-plane antiferromagnetic couplings between the spin of Ni $d_{x^2-y^2}/d_{z^2}$ orbitals,
\begin{eqnarray}
H_{J}=\sum_{\langle ij\rangle\alpha}J^\alpha_{ij}(\mathbf{S}_{i\alpha}\mathbf{S}_{j\alpha}-\frac{1}{4}n_{i\alpha}n_{j\alpha})
\end{eqnarray}
where
$\bm{S}_{i\alpha}=\frac{1}{2}c_{i\alpha\sigma}^{\dagger}\bm{\sigma}_{\sigma\sigma'}c_{i\alpha\sigma'}$
is the local spin operator and $n_{i\alpha}$is the local density
operator for Ni $\alpha$ orbital ($\alpha=1,2$). $\langle ij\rangle$ denotes the in-plane and out-of-plane nearest neighbors. and the in-plane and out-of-plane couplings are $J^{\alpha}_{x/y}=J_\alpha$ and $J^{\alpha}_{z}=J'_\alpha$, respectively.   By performing the Fourier transformation, $H_{J}$ in momentum space reads
\begin{eqnarray}
H_{J}&=&\sum_{\eta\alpha\mathbf{k}\bm{k}'}V^\alpha_{\mathbf{k},\mathbf{k}'}P^\dag_{\eta\alpha }(\bm{k})P_{\eta\alpha}(\bm{k}')
+\sum_{\bm{k}\bm{k}'}W^{\alpha}_{\bm{k},\bm{k}'}B^\dag_{\alpha}(\bm{k})B_{\alpha}(\bm{k}'),\nonumber\\
\end{eqnarray}
with the intra-layer pair operator $P^\dag_{\eta\alpha }(\bm{k})=c_{\eta\alpha\uparrow}^{\dagger}(\bm{k})c_{\eta\alpha\downarrow}^{\dagger}(-\bm{k})$ and the inter-layer pair operator $B^\dag_{\alpha}(\bm{k})=c_{t\alpha\uparrow}^{\dagger}(\bm{k})c_{b\alpha\downarrow}^{\dagger}(-\bm{k})+c_{b\alpha\uparrow}^{\dagger}(\bm{k})c_{t\alpha\downarrow}^{\dagger}(-\bm{k})$. Here $V^\alpha_{\mathbf{k},\mathbf{k}'}=-\frac{2J_\alpha}{N}\sum_{\pm}(cosk_x\pm cosk_y)(cosk'_x\pm cosk'_y)$ and $W^\alpha_{\mathbf{k},\mathbf{k}'}=-\frac{J'_\alpha}{2N}$ and $N$ being the number of lattice sites. We investigate the pairing state for both undoped and doped systems and neglect the the no-double-occupance constraint on this t-J model and perform a mean-field decoupling.
With this, the total Hamiltonian can be written as,
\begin{eqnarray} H_{MF}&=&\sum_{\mathbf{k}}\Psi_{\mathbf{k}}^{\dagger}A(\mathbf{k})\Psi_{\mathbf{k}}+\frac{N}{2}\sum_{\alpha,\nu=s,d}\frac{|\Delta^\alpha_{\nu}|^2}{J_\alpha} \nonumber\\
&+&2N\sum_{\alpha}\frac{|\Delta^\alpha_{inter}|^2}{J'_\alpha},\\
A(\mathbf{k})&=&\left(\begin{array}{cc}
h(\mathbf{k}) & \Delta_{\uparrow\downarrow}(\mathbf{k}) \\
 \Delta^{\dagger}_{\uparrow\downarrow}(\mathbf{k}) & -h^{*}(-\mathbf{k}) \\
 \end{array}\right), \nonumber\\
 \Delta_{\uparrow\downarrow}(\mathbf{k})&=&\left(\begin{array}{cccc}
\Delta^t_{1}(\mathbf{k}) &  & \Delta^{tb}_{1} &  \\
  & \Delta^t_{2}(\mathbf{k}) & & \Delta^{tb}_{2}\\
  \Delta^{bt}_{1}&   &  \Delta^b_{1}(\mathbf{k})  & \\
   &   \Delta^{bt}_{2}&   & \Delta^b_{2}(\mathbf{k}) \\
 \end{array}\right),
\end{eqnarray}
where $\Psi_{\mathbf{k}}^{\dagger}=(\psi^\dag_{\bm{k}\uparrow},\psi^T_{-\bm{k}\downarrow})$, $\Delta^{t/b}_{\alpha}(\mathbf{k})  =  \Delta^{\alpha,t/b}_{s}(cosk_x+cosk_y)+\Delta^{\alpha,t/b}_{d}(cosk_x-cosk_y)$, $\Delta^{tb/bt}_\alpha=\Delta^\alpha_{inter}$, and
\begin{eqnarray}
\Delta^{\alpha,\eta}_{s/d} &=& -\frac{2J_\alpha}{N}\sum_{\mathbf{k}'}d^{\alpha,\eta}_{\mathbf{k}'\uparrow}(cosk'_x\pm cosk'_y),\\
\Delta^\alpha_{inter} &=& -\frac{J'_\alpha}{2N}\sum_{\mathbf{k}'}g^\alpha_{\mathbf{k}'\uparrow},
\end{eqnarray}
with $d^{\alpha,\eta}_{\mathbf{k}'\uparrow}=\left\langle
c_{\eta \alpha\downarrow}(-\mathbf{k}')c_{\eta \alpha \uparrow}(\mathbf{k}')\right\rangle
$ and $g^\alpha_{\mathbf{k}'\uparrow}=\left\langle c_{b\alpha\downarrow}(-\bm{k}) c_{t\alpha\uparrow}(\bm{k})+c_{t\alpha\downarrow}(-\bm{k})c_{b\alpha\uparrow}(\bm{k})\right\rangle$.. $A(\mathbf{k})$ can be diagonalized by an unitary transformation
$U_{\mathbf{k}}$ with and the Bogoliubov quasiparticle eigenvalues $E_{m+4}=-E_{m}$ with $m=1,2,3,4$. The self-consistent gap equations are
\begin{eqnarray}
\Delta^{1,t/b}_{s/d} & = & -\frac{2J_1}{N}\sum_{\mathbf{k},m}(cosk_x\pm cosk_y)U_{5/7,m}^{*}(\mathbf{k})U_{1/3,m}(\mathbf{k})F[E_{m}(\mathbf{k})]\nonumber\\
\Delta^{2,t/b}_{s/d} & = & -\frac{2J_2}{N}\sum_{\mathbf{k},m}(cosk_x\pm cosk_y)U_{6/8,m}^{*}(\mathbf{k})U_{2/4,m}(\mathbf{k})F[E_{m}(\mathbf{k})]\nonumber\\
\Delta^1_{inter} & = & -\frac{2J'_1}{N}\sum_{\mathbf{k},m} [U_{7,m}^{*}(\mathbf{k})U_{1,m}(\mathbf{k})\nonumber\\
&&+U_{5,m}^{*}(\mathbf{k})U_{3,m}(\mathbf{k})]F[E_{m}(\mathbf{k})]\nonumber\\
\Delta^2_{inter} & = & -\frac{2J'_2}{N}\sum_{\mathbf{k},m} [U_{8,m}^{*}(\mathbf{k})U_{2,m}(\mathbf{k})\nonumber\\
&&+U_{6,m}^{*}(\mathbf{k})U_{4,m}(\mathbf{k})]F[E_{m}(\mathbf{k})]
\end{eqnarray}
where $F[E]$ is Fermi-Dirac distribution function, $F[E]=1/(1+e^{E/k_{B}T})$. The above equations can be solved self-consistently, varying the doping and the value of $J_1,J_2,J'_1,J'_2$.

\end{document}